\documentclass{PoS}

\title{First Results with HIJING++ on High-energy Heavy Ion Collisions}

\ShortTitle{HIJING++}

\author{\speaker{G\'abor Papp}$^{a}$, Gergely G\'abor Barnaf\"oldi$^{b}$, G\'abor B\'\i r\'o$^{a,b}$,
        Miklos Gyulassy$^{b,c,d,e}$, Szilveszter Harangoz\'o$^{b}$, Guoyang Ma$^{c}$, P\'eter L\'evai$^{b}$, Xin-Nian Wang$^{c,d}$, 
        Ben-Wei Zhang$^{c}$ \\
        \llap{$^{a}$}Institute for Physics, E\"otv\"os Lor\'and University, 1/A P\'azm\'any P. S\'et\'any, H-1117, Budapest, Hungary\\
        \llap{$^{b}$}Wigner Research Centre for Physics of the Hungarian Academy of Sciences, 29-33 Konkoly-Thege Mikl\'os Str, H-1121 Budapest, Hungary\\
        \llap{$^{c}$}Institute of Particle Physics, Central China Normal University, Wuhan 430079, China\\
        \llap{$^{d}$}Nuclear Science Division, MS 70R0319, Lawrence Berkeley National Laboratory, Berkeley, California 94720 USA \\
        \llap{$^{e}$}Pupin Lab MS-5202, Department of Physics, Columbia University, New York, NY 10027, USA
        E-mail: \email{pg@elte.hu}}

\abstract{We present preliminary results with HIJING++ (3.1.1) for identified hadron production in high-energy heavy ion collisions at LHC energies. The recently developed HIJING++ version is based on the latest version of PYTHIA8 and contains all the nuclear effects that have been included in the HIJING2.552, which will be improved by a new version of the shadowing parametrization and jet quenching module. Here, we summarize the structure and the speed gain due to parallelization of the new program code, also presenting some comparison between experimental data.}

\FullConference{
12th International Workshop on High-pT Physics in the RHIC/LHC Era\\
		2-5 October, 2017\\
		University of Bergen, Bergen, Norway}

\begin{document}
\section{Introduction}
\label{sec:intro}

The original HIJING~\cite{HIJING} (Heavy Ion Jet INteraction Generator) Monte Carlo model was developed by M. Gyulassy and X.-N. Wang with special emphasis on the role of minijets in proton-proton (pp), proton-nucleus (pA) and nucleus-nucleus (AA) reactions at collider energies in a wide range from 5 GeV to 2 TeV. The original program itself is written in FORTRAN, and is based on the FORTRAN version of PYTHIA (version 5)~\cite{PYTHIAv5}, ARIADNE~\cite{ARIADNE}, and the CERNLIB package PDFLIB~\cite{CERNLIB}. The program is a widely used particle event generator, however is not parallel and lacks the portability to experimental platforms. 

HIJING was developed to incorporate nuclear effects, such as collision geometry, shadowing, Cronin effect, jet quenching, multiple collisions. PYTHIA is used for hard jet production with additional shadowing and transverse momentum exchange. Soft beam jets are modeled by diquark-quark strings with gluon kinks along the lines of the Lund FRITIOF~\cite{FRITIOF} and dual parton model (DPM). In addition, multiple low-$p_T$ exchanges among the endpoint constituents are included to model initial state interactions. The excited strings decay according to the ARIADNE program code, which will be upgraded in the future versions with Gunion-Bertsch radiation\cite{Gunion-Bertsch} .

\section{Main objectives}
\label{sec:objects}

Since the release of the first HIJING version a number of underlying libraries has undergone a major upgrade connected with structural changes, and they got rewritten to C++, becoming a standard now in the high-energy community. Experimental platforms like AliRoot~\cite{ALIROOT} are based on that programing language. Furthermore, with the new releases of C++ the parallelization of codes in multithreaded environment became much easier.

Hence, we decided to upgrade HIJING accordingly, to be a genuine, C++ based, modular event generator, with the most recent versions of PYTHIA8~\cite{PYTHIA} and LHAPDF6~\cite{LHAPDF}, and be compatible with AliRoot. The original shadowing functions was known to produce overshadowing at RHIC energies, so it was extended to have $Q^2$ dependence based on the HOPPET~\cite{HOPPET,USHOPPET} code. Indeed, that reduced somewhat the shadowing at RHIC energies, however, made no change at LHC energies, so further work is needed here.

\section{The HIJING++ program}
\label{sec:hijingcode}

Since one of the basic components of HIJING, the PYTHIA event generator switched to C++ in 2006, it was natural to make a similar change with HIJING, and use the classes already introduced be PYTHIA8, extending them appropriately for HIJING. In HIJING, we took great care of parallelization, which was not an easy task, because PYTHIA8 code was not written with such intention. Fortunately, the structure of PYTHIA8 was still quite flexible to introduce the necessary changes, and even for example there is a possibility to replace PYTHIA's pseudo-random number generator (PRNG) by a faster GPU-based one~\cite{Barnafoldi-NagyEgri}. 

\begin{figure}[h]
\begin{center}
\includegraphics[width=0.95\linewidth, height=95mm]{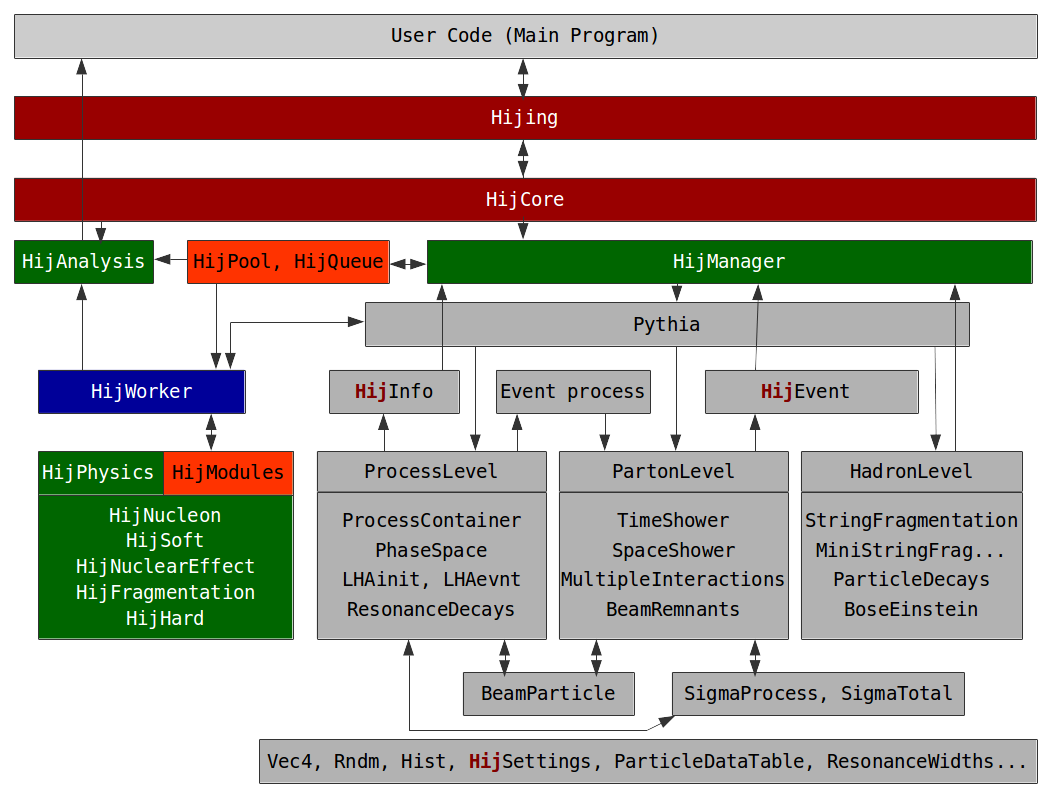}
\caption{The structure of the HIJING++ 3.1.1 program code.}
\label{fig:hijing-struc}
\end{center}
\end{figure}

The main structure of the code is presented in Fig.~\ref{fig:hijing-struc}, where colored boxes represent the newly-included {\tt Hijing} modules and modifications neglecting cross-links. The {\tt Hijing} class contains all the physics were coded in the FORTRAN subroutines, based on the latest version of HIJING version 2.552~\cite{HIJING2}. Due to the object oriented being of the C++, the original structure was optimized for modularity and  compiler's  improved parallel supports. The high-energy nuclear physics related part (hard collisions, soft collisions, fragmentation, Cronin, jet quenching) are moved to the {\tt HijPhysics} class, where they can be called modularly, with the possibility to alter them to user supplied modules. The {\tt HijQueue} and {\tt HijManager} classes are responsible for distributing the separate Hijing events in a parallel environment. So far only threadbase parallelization is supported in version 3.1.1, however, a distributed parallel environment, such as MPI can be easily implemented.

A useful new feature is the {\tt HijAnalysis} class, being an interface to user defined data collections within the run. The code has several entry points for user provided and built-in histogram collection modules, which are configurable in the user (main) code with one line commands.

\section{Status of the project and results}
\label{sec:results}

The new version of C++ code, HIJING++ 3.1.1 is currently under testing, before release. The physical models are still based on the ones of version 2.552 (FORTRAN) with changes in hard sector due to PYTHIA8, and modification of the Cronin scattering.

\subsection{HIJING++ results in proton-proton collisions}
\label{sec:hijing-pp}

\begin{figure}[h]
\begin{center}
\includegraphics[width=1.0\linewidth, height=75mm]{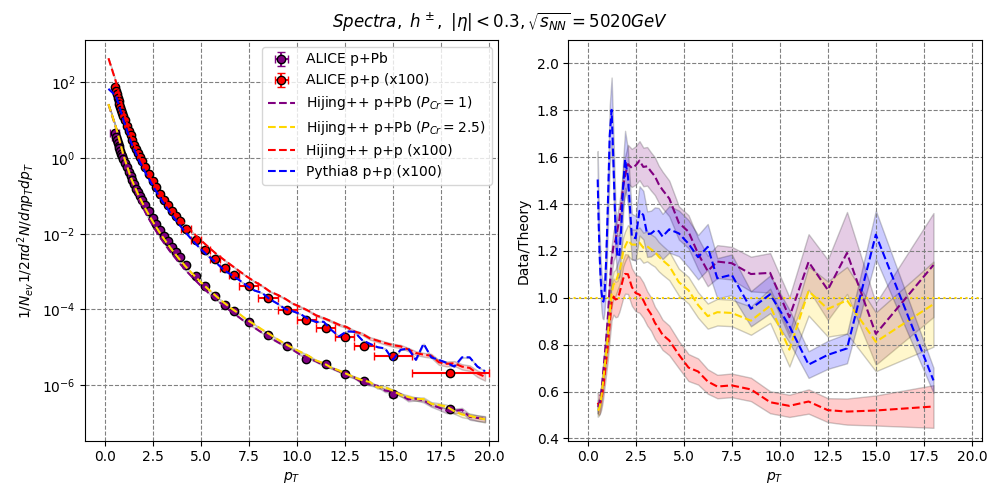}  \vspace*{-1cm}
\caption{{\bf Left panel}: HIJING++ calculation for pp$\to$ h${}^\pm$  at 5020 GeV yield $1/N_{ev} d^2N/dp_T^2d\eta$ in the midrapidity $|\eta| < 0.3$ (dashed red line) compared to the interpolated experimental ``data'' (red circle) and PYTHIA8 (blue purple line) calculations. Spectra were shifted for better visibility. With purple dashed line the HIJING++ yield is presented. {\bf Right panel}: Data/Theory ratio for pp and pPb reactions, with same color coding as for the left panel.}
\label{fig:pp-spectra}
\end{center}
\end{figure}

In Figure~\ref{fig:pp-spectra} we plotted the charged hadron yields at LHC $\sqrt{s}=$5020 GeV energy, midrapidity $|\eta| < 0.3$ window, calculated for 100 million events in the present HIJING model and compared to interpolated experimental ``data''~\cite{ALICE5020} and 85 million events from PYTHIA8 calculation. Since the hard scattering in HIJING++ is based on PYTHIA8, the different behaviour at mid and high $p_T$ is related to the different tunes used, HIJING uses the default settings (with no multi particle interaction), while the PYTHIA8 calculation was performed with the Monash tune optimized for LHC energies.

The pp fit is within 50\% both for HIJING++ and PYTHIA8, the former performing better at mid $p_T$, while the latter at high $p_T$. The agreement is fair, however, some fine tuning is required to increase the accuracy, like following the Monash-tune to add intrinsic $k_T$ width, and change in the hard scale. We note, HIJING++ calculations have not been tuned yet, we applied settings for PYTHIA8 adopted from the old PYTHIA verion 5. 

\subsection{HIJING++ results in proton-nucleus collisions}
\label{sec:hijing-pA}

We tested the HIJING++ code also in minimum bias proton-lead (pPb) collision at LHC energy, $\sqrt{s_{NN}}=5.02$ TeV, calculating the yield of charged hadrons at midrapidity,  $|\eta| < 0.3$. For this case 20 million events were generated with impact parameter dependent shadowing, and Cronin multiple scattering with average transverse momentum square exchange of 4 GeV${}^2/c^2$. The hard separation scale was set to 5.05 GeV/$c$.

This code already contains an improved version of the final transverse momentum exchange (Cronin effect), introducing an energy dependent width for the distribution of the average transverse momentum exchange square~\cite{kTpQCD}. The old version had a very weak dependence for the width, working at lower energies, however, underestimating the effect at LHC energies.

\begin{figure}[h]
\begin{center}
\includegraphics[width=\linewidth,height=70mm]{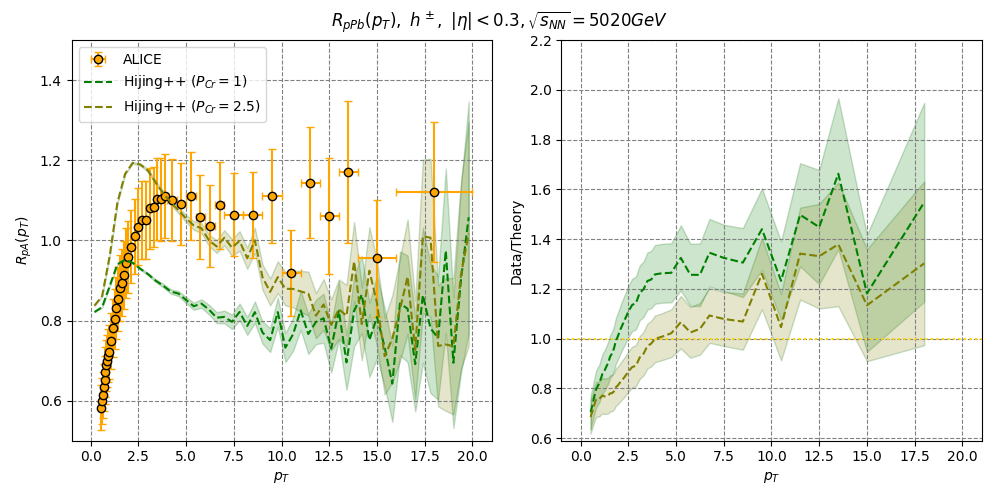}  \vspace*{-1cm}
\caption{Nuclear modification factor, $R_{pPb}$ for $|\eta| < 0.3$ in pPb collisions at $\sqrt{s_{NN}}=5.02$ TeV in comparison to the ALICE data from Ref~\cite{ALICE5020} (left panel), and Data/Theory (right panel).}
\label{fig:RpPb}
\end{center}
\end{figure}

In Figure~\ref{fig:pp-spectra} the results obtained by present HIJING++ on hadron spectra are presented in pPb collisions at $\sqrt{s_{NN}}=5.02$ TeV energy in comparison with ALICE~\cite{ALICE5020} data. The pPb HIJING++ simulations are also within 30\% agreement with the data . 

In Figure~\ref{fig:RpPb} we present the nuclear modification factor calculation with HIJING++ contrasted to the experimental result~\cite{ALICE5020}. There is general suppression in the calculation which was just recently pinned down to be the effect of pushing too much away from the independent fragmentation idea, and hence, pushing the original $N_{bin}$ scaling in the direction of $N_{part} < N_{bin}$. We are currently focusing on this issue, and believe, that with the proper treatment we achieve a much better agreement with the data.

\subsection{Parallelization}
\begin{figure}[h]
\begin{center}
\includegraphics[width=85mm,height=55mm]{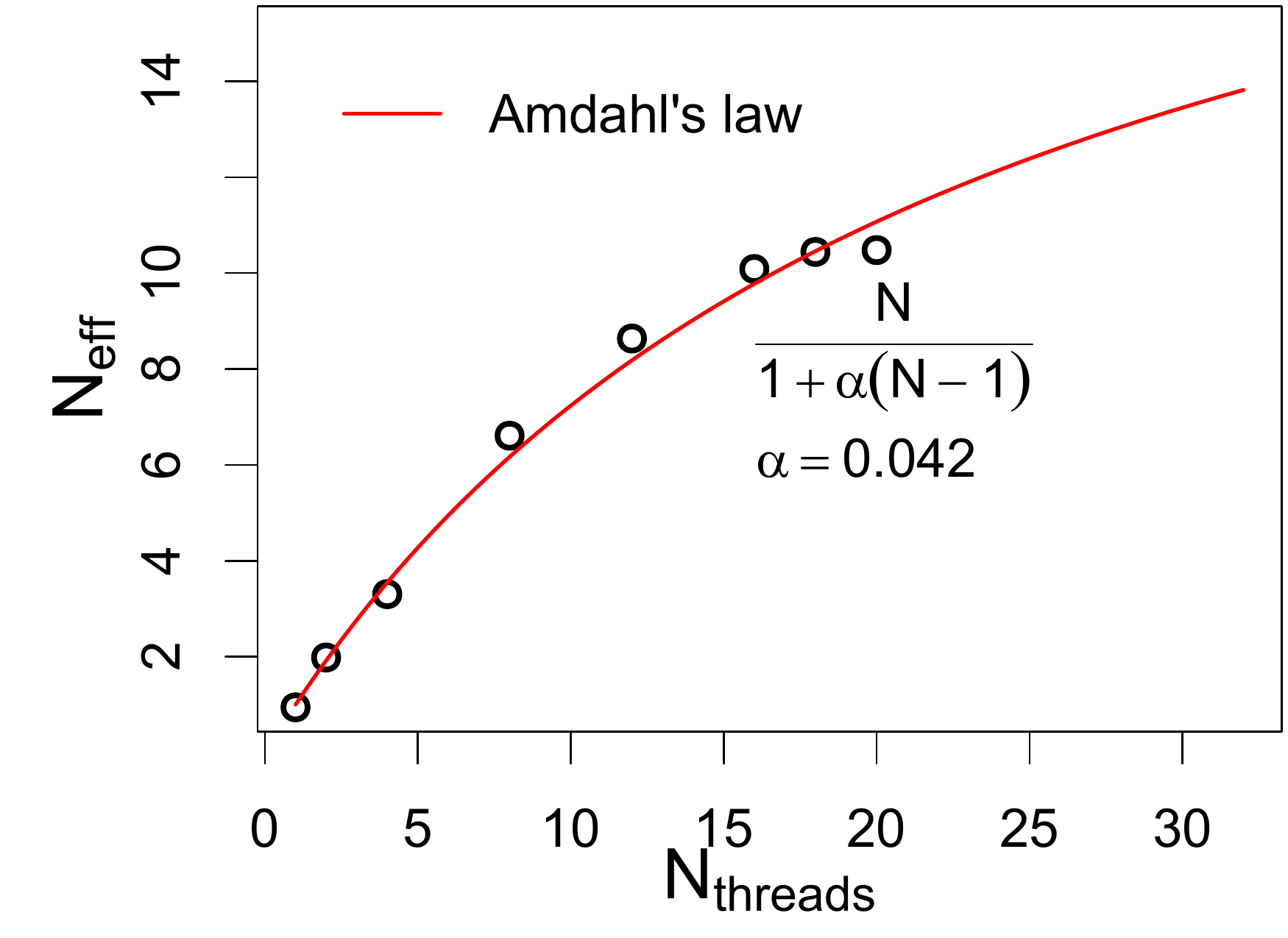} 
\caption{Amdahl's law fit of calculation speed on the number of threads. According to the fit the non-parallel part of the code is 4.2\%.}
\label{fig:Amdahl}
\end{center}
\end{figure}
\label{sec:hijing-parallel}
In Figure~\ref{fig:Amdahl} we present the speedup in computational time in parallel usage, compared to single-thread performance. The tests were performed on a 20 thread Intel Xeon E312 processor virtual machine at Wigner Datacenter. The measured points follow nicely Amdahl's law~\cite{AMDAHL}, with a fit on the non-parallel portion of the code (mostly queue management) being 4.2\%, giving the maximal speedup to be around 24. We believe, that this speedup can be still improved by tuning the queue sizes and left for further work.

\section{Ongoing developments and future perspectives}
\label{sec:devels}

In parallel to the test phase of the coding, a sophisticated theoretical development of the model is ongoing.
This is a challenging task, since in these phenomenological models we need to satisfy all previous and recently measured data in several manners.
 
Here we list some of these updates and features, which will be present in the final, public code: 
\begin{itemize}
\item The original HIJING shadowing function, and also its modified, $Q^2$ dependent version are producing too much shadowing, so we are considereing to implement other shadowing models.

\item A contemporary jet energy loss module is under development. This module is enable users to include various jet-quenching models in the future, such as we include Gyulassy\,--\,L\'evai\,--\,Vitev (GLV)~\cite{GLV}.  

\item The run-time of the HIJING++ is comparable to the previous, FORTRAN based version, however, this new version is suitable for parallel architectures, which may speed up the calculations considerably. The present code structure is capable to provide thread-level and/or MPI parallelization.  With increasing energy it was also necessary to change from single precision to double one.

\item Since the new version of the code is written in the modular C++, it is natural to include it to huge detector simulation frameworks, like ALICE's AliRoot~\cite{ALIROOT}. Due to the modularity, merging of these simulation frameworks can reduce their volume and the number of cross links, thus simulations become more memory consumable.  

\item Since the code is written completely in C++ a (partial) parallel-platforms supported version of the program is planned to develop. The preliminary tests showed, that changing the random number generator to a GPU-based version, can result in a slight increase of the speed. Optimalization for various multi-core and parallel architectures is on the wish list. 
\end{itemize}

\section{Summary and outlook}
\label{sec:summary}

Authors summarize here the results calculated by the 3.1.1 pre-release version of the Monte Carlo heavy ion jet interaction generator, HIJING++. Adopting the structure of the PYTHIA8, the {\tt Hijing} class were invented, including the nuclear mechanism modeled in the original (FORTRAN) HIJING 2.5x version. We presented the comparison of charged hadron yields in proton-proton and proton-lead collisions at LHC $\sqrt{s_{NN}}=5.02$ TeV energy to experimental data. It was shown, that the code is highly parallel, with approximately 4\% on non-parallelized portion.

Present milestone aims to communicate the status of this software development, moreover, give perspectives for the forthcoming applicabilities and features of the soon-to-be-released open source HIJING++ for the next generation of heavy-ion collision measurement, simulations, and facilities at future colliders.

\section*{Acknowledgements}

This work was supported by the Hungarian-Chinese cooperation grant No T\'eT 12 CN-1-2012-0016 and No.
MOST 2014DFG02050, Hungarian National Research Fund (OTKA) grant K120660 and THOR COST action 15213. We acknowledge the support of the Wigner GPU laboratory, and Wigner Datacenter. G. B\'\i r\'o was supported by the \'UNKP-17-3 New National Excellence Program of the  Ministry of Human Capacities.


\begin{thebibliography}{99}
\bibitem{HIJING} X.N. Wang, M. Gyulassy,  Phys. Rev. {\bf D44}, 3501  (1991).

\bibitem{HIJING2} W.T. Deng, X.N. Wang, R. Xu, Phys. Rev. {\bf C83}, 014915 (2011).

\bibitem{PYTHIA} T. Sj\"ostrand, Comput. Phys. Commun. {\bf 191}, 159 (2015).

\bibitem{ALIROOT} AliRoot: http://aliweb.cern.ch/Offline/AliRoot/Manual.html (2017)

\bibitem{PYTHIAv5} T.~Sjostrand,
  Comput.\ Phys.\ Commun.\  {\bf 82}, 74 (1994).

\bibitem{ARIADNE} L. L\"onnblad, Comput. Phys. Comm. {\bf 71}, 15 (1992).

\bibitem{CERNLIB} CERNLib:  https://cernlib.web.cern.ch/cernlib/ (2017)

\bibitem{FRITIOF}  B.~Nilsson-Almqvist and E.~Stenlund,
  Comput.\ Phys.\ Commun.\  {\bf 43}, 387 (1987).

  
\bibitem{LHAPDF} A. Buckley, J. Ferrando, S. Lloyd, K. Nordstr\"om, B. Page, M. R\"ufenacht, M. Sch\"onherr, G. Watt,  Eur.Phys.J. {\bf C75} 3, 132 (2015).

\bibitem{HIJINGsh} X.N. Wang, Phys. Rev.{\bf C61}, 064910 (2001).

\bibitem{DIPSY} C. Flensburg,  Prog.Theor.Phys.Suppl. {\bf 193}, 172 (2012). 

\bibitem{Gunion-Bertsch} J.F. Gunion and G. Bertsch, Phys. Rev.{\bf D25}, 746 (1982)

\bibitem{GLV} M.Gyulassy, P.Levai, I.Vitev, Phys .Rev. Lett. {\bf 85}, 5535 (2000).

\bibitem{HOPPET} A. Vogt, S. Moch, J.A.M. Vermaseren, Nucl. Phys. {\bf B691}, 129 (2004) 

\bibitem{USHOPPET}G. Ma, G.G. Barnaf\"oldi, Weitian Deng, Sz. Harangoz\'o, G. Papp, X-N. Wang, B-W. Zhang, DGLAP-evolved Shadowing Parametrization for Simulating High-energy Nucleus-Nucleus Collisions in HIJING, ({\sl in preparation})

\bibitem{Barnafoldi-NagyEgri} G.G. Barnaf\"oldi, M.F. Nagy-Egri, GPU-based PRNG for Monte Carlo Particle Event Genarators ({\sl in preparation})

\bibitem{ALICE5020} ALICE Collaboration. {\em Phys. Rev. Letters}  {\bf 2013}, {\em 110}, 082302.







\bibitem{kTpQCD}  Y. Zhang, G. Fai, G. Papp, G.G. Barnafoldi, P. Levai, {\em Phys. Rev.} {\bf 2002}, {\em C65}, 034903.

\bibitem{AMDAHL} G.M. Amdahl, {\em AFIPS Conference Proceedings} {\em 30}, 483.

\end{thebibliography}
\end{document}